# NONLINEAR NANO-OPTICS AND ITS APLICATION: BOND MODEL FOR SEMICONDUCTOR CHARACTERIZATION

**Hendradi Hardhienata**[1,2]

[1]Center for surface- and nanoanalytics (ZONA), Johannes Kepler University, Altenbergerstr. 69, 4040 Linz, Austria
[2] Theoretical Physics Division, Department of Physics, Bogor Agricultural University, Jl. Meranti, Gedung Wing S, Kampus IPB Darmaga, Bogor 16680, Jawa Barat, Indonesia
Email: hendradi_h@yahoo.com

## ABSTRACT

This paper begins with a brief history of nonlinear optics and how the bond model emerges as a need to obtain a better physical picture of nonlinearity. A description about the possible application of the bond model is presented afterward followed by a theoretical explanation of its fundamentals. The possibility to develop this model for future application such as surface reconstruction characterization, real time surface biosensor monitoring, and semiconductor surface bandgap determination is also presented at the end.

## 1. INTRODUCTION

Our world is nonlinear in the sense that the response of physical systems towards external perturbation is in general nonlinear. The nonlinearity usually becomes significant when the perturbation is large e.g. in nonlinear optics it is often in the form of an intense laser field. Although the history of nonlinear optics can be traced back to 1870 when John Kerr demonstrated that the refracted index of certain solids and liquids can be slightly altered if exposed to a strong DC field [1], it is not until the discovery of the laser by Townes [2] in 1955 and Maiman [3] in 1960 that research in nonlinear optics began to flourish. The first laser generated nonlinear phenomena was the discovery of second harmonic generation (SHG) by Franken in 1961 [3]. The second harmonic field on the detector was so small so that it was regarded as a dirt spot by the editor of Physical Review Letter and was erased from the paper. Nevertheless, it was soon found that the higher harmonic fields generated inside the nonlinear materials can be greatly enhanced using phase matching. Since then, nonlinear phenomena was more than just a theoretical curiosity. Afterwards, many other nonlinear phenomena were discovered, these are intensity dependent refractive index [4], Third Harmonic Generation (THG) [5], and Electric Field Enhanced Second Harmonic Generation (EFISH) [6].

It is remarkable that already at such an early stage since the discovery of SHG that a firm theoretical framework was provided by the group of Pershan [7,8]. They also developed a theoretical basis to the generation of nonlinearity from crystal surfaces particularly that have the property of parity or inversion symmetry [6]. A more explicit analysis regarding nonlinear generation from such surfaces was performed using phenomenology models applying polarisable sheet models [9,10] which was able to reproduce experimental data but unfortunately involves a large number of fitting parameter which blurs the physics. Recently, the Simplified Bond Hyperpolarizability Model (SBHM) [11] was developed by Aspnes group to address these shortcomings. Although SBHM is not an ab initio model as the hyperpolarizabilities must be fitted with experimental data it has the advantage over the previous phenomenological model because it requires less independent fitting parameter and due to the clarity of the physics used in deriving this model.

The bond model is based on the classical notion that charges in a nonlinear material can be regarded as localized particles. It can be classified as a nano optical model because it gives a microscopic picture down to the nano scale regarding the interaction to between the driven charge and the driving field. In fact, it has been possible to investigate molecular bond orientation in $SiO_2$ semiconductor interfaces using this model which successfully reproduced experiment result for a nonvicinal Si(111) orientation [12]. Moreover, it was also showed by Kwon *et. al*.13] that the nonlinear susceptibility obtained from SBHM is in agreement with causality, e.g. Kramers-Krönig relation.

## 2. ADVANTAGES OF THE NONLINEAR BOND MODEL

The advantages of using nonlinear optical techniques, even though it requires a high intensity short pulse laser, are as follows: it is a real time and accurate method in extracting information from a nonlinear material. In addition, the experiment can be performed at non vacuum under room temperature condition and is non destructive, if treated properly. Although previous authors of the bond model focused merely on the 'physics' by investigating the nonlinear susceptibilities, its potential application to study the atomic orientation was straightforward since the radiation in this model was obtained by assuming that the nonlinear fields were generated from charges that are moving only along the atomic bonds. By using rotational anisotropy spectroscopic (RAS) method and femto/pico second lasers it was shown that the molecular orientation of a $SiO_2$ surface can be found using the bond model [12]. Such information can be important if one wants to investigate the surface properties of a material which is of high demand nowadays due to the widespread application of thin film technology. Relevant application of the model include the possibility to detect surface properties of metals that undergo corrosion process [14] or to detect the vicinality (e.g. slope or terraces) of a crystal [11,13]. Other possible application would be in the field of electrical engineering particularly in enhancing the understanding of semiconductor devices such as band gap analysis using a DC electric field which break the parity symmetry [15]. We recently show that such symmetry breaking can be modelled using the bond model for the case of metal oxide semiconductors (MOS) [16].

The reason why nonlinear optical techniques are able to characterize surface properties is based to the fact that even higher harmonic modes generated by dipole radiation inside the bulk of a material that exhibit parity symmetry or invariant under spatial coordinate inversion are forbidden [6]. It can be shown using group theory that these materials gives a zero even order nonlinear susceptibility. A straightforward consequence is that the even order nonlinear polarization is zero, hence the even order nonlinear intensity must also be zero. For example, the second order nonlinear polarization $P_j^{(2)}(2\omega)$ is given by:

$$P_j^{(2)}(2\omega) = \chi_{jkl}^{(2)} E_k(\omega) E_l(\omega) \tag{1}$$

where $\chi_{jkl}^{(2)}$ is the second order nonlinear susceptibility which only depends on the crystal symmetry and $E_k(\omega)$ is the applied input field. For a Silicon (Si) semiconductor belonging to the $O_h$ point group, the second rank tensor can be obtained via [17,18]:

$$s_{ijk}'(\phi) = R_{im} R_{jn} R_{ko} s_{mno}(\phi) \tag{2}$$

here $s_{ijk}'$ is the second order susceptibility in the new coordinate frame and $R$ is the matrix that transforms the previous coordinate system into the new one with regard to a specific symmetry operation e.g. can be a rotational, translational, mirror matrix, *etc*. When Eq. (2) is evaluated for all the allowed symmetry operation including parity the second harmonic susceptibility is zero which is in agreement with standard group theoretical literature [17,18] thus no SHG signal can occur from the bulk.. However, at the surface, the parity symmetry is broken due to anisotropy and absence of an upper layer. Therefore contrary to bulk, the even order nonlinear dipoles can now radiate SHG.

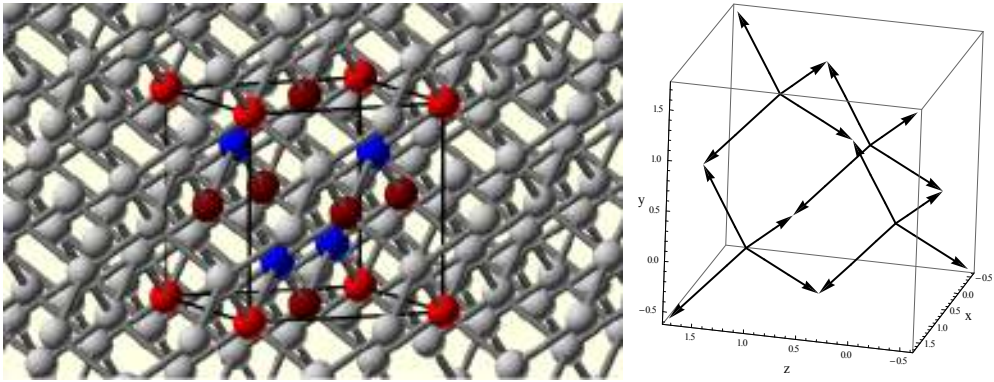

Figure 1. Bulk and surface crystal orientation of a tetrahedral Si(001) facet [19] (left). Bond vector definition for the corresponding crystal orientation (right).

Consider a Si(001) surface as given by Fig. 1. The $O_h$ symmetry of the surface is broken to $C_{2v}$ and the second oder nonlinear susceptibility is not zero anymore and takes the form [20]:

$$\chi^{(2)}{}_{ijk} = \begin{pmatrix} 0 & 0 & s_{131} \\ 0 & 0 & 0 \\ s_{131} & 0 & 0 \\ 0 & 0 & 0 \\ 0 & 0 & s_{232} \\ 0 & s_{232} & 0 \\ s_{311} & 0 & 0 \\ 0 & s_{322} & 0 \\ 0 & 0 & s_{333} \end{pmatrix} \qquad (3)$$

Eq. (3) shows that there are at least five independent tensor elements which can further be reduced to three by applying intrinsic permutation and Kleinman symmetry (see Eq. 6) [20]. Therefore SHG is a popular technique to investigate surface properties of centrosymmetric semiconductors such as Si. It has to be stated however that other nonlinear sources can also occur from the bulk such as quadrupole contribution and spatial dispersion (decay of the field). It has also been reported that bulk contribution can be small in certain cases such as in Si [21] although determining the ratio of surface and bulk contribution in several facet orientation such as Si(111) remains difficult and is an active area of research.

## 3. BOND MODEL FUNDAMENTALS

The bond model was inspired by the fact that one can obtain both the linear and nonlinear optical far field by summing over all dipoles. This approach was developed for linear optics by Ewald [22] and Oseen [23]. For a normal incidence laser field (Fig. 2) the total field produced by all dipoles is given by an integral equation [24]:

$$E_T(z) = E_i^{(0)} e^{ik_0 z} + \frac{ik_0}{2}[\varepsilon(\omega)-1]\int_0^\infty dz' E_T(z') e^{ik_0|z-z'|} \qquad (4)$$

where $E_i^{(0)}$ is the incoming field and $E_T(z)$ is the total field. Here $\varepsilon(\omega)$ is the dielectric permittivity and $k_0$ is the incoming wave vector. This approach was later extended by Aspnes and Adles into nonlinear optics which surprisingly turned out to be more simple due to the absence of the incident nonlinear field (because it is generated inside the material) thus eliminating the need in solving a self consistent equation [25].

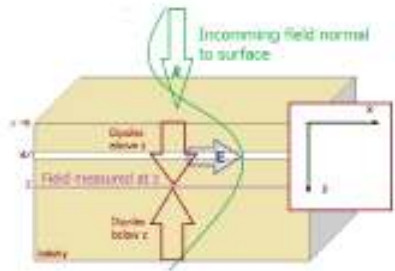

Figure 2. Summing over all dipoles for a incomming field at normal incidence. SBHM extends this principle to nonlinear optics with the addition that the motion is only along the bonds

An additional assumption of the bond model in its earliest version is that the dipoles are oscillating both harmonically and anharmonically only paralel along the bonds. This is justified by quantum mechanical models (See Fig. 3 *left*) using linear combination of atomic orbital (LCAO) calculation such as tetrahedral crystals (sp3 orbitals) where the electron probability wave function are highest along the bonds or from density functional theory (DFT) (FIG. 3 *right*) where the electron density for such simple crystal system is highest along the bonds.

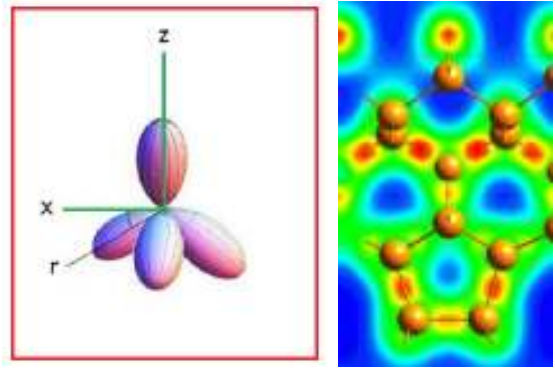

Figure 3. Sp3 model of a Si(111) tetrahedral facet (left). DFT calculation showing high electron density along the bonds, figure Taken from MPI Halle group (Ref. [26]) (right).

Another way to look at it is classically, namely assuming that the bond are rotationally symmetric for SHG driven charges so that it can be modelled as a one dimensional lorentz model whose potential is oriented mainly paralel to the bonds (FIG 4).

The second order susceptibility in SBHM is given by [11]:

$$\chi^{(2)}_{jkl} = \frac{\alpha_2}{V} \sum_{j=1}^{4} b_j(\phi) \otimes b_j(\phi) \otimes b_j(\phi) \tag{5}$$

Where $b_j(\phi)$ is the defined bond directions as a function of the azhimut angle $\phi$ (the angle around the z axis), $\alpha_2$ is the SHG hyperpolarizability and $V$ is the volume of the unit cell which is just a scaling factor. Thus the bond model takes a rather physical view in obtaining an expression for the susceptibility: by taking granted the motion of charges along the bonds rather than previous phenomenolgical models which obtain the susceptibiliy from the crystal point group. Therefore in SBHM, the symmetry is already represented by the bond direction in Eq. 5.

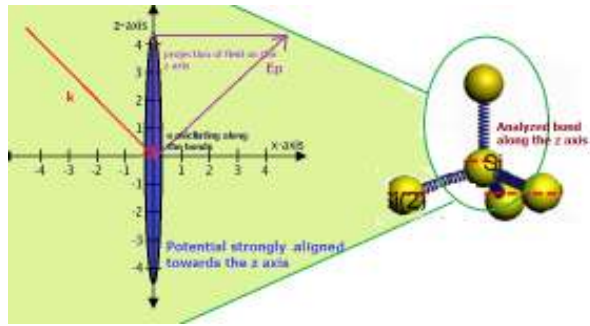

Figure 4. The potential is assumed to be alligned paralel along the bonds which is valid for Second Harmonic driven fields. When a driving field produces charge radiation it will oscillate strongest if the field is aligned paralel to the atomic potential. Here only the potential along the *z*-axis (*up* bond) is shown.

The next question is whether such approach although physically meaningful is also a valid from the viewpoint of crystal symmetry. In other words, are the nonlinear susceptibility obatined from SBHM in agreement with group theory? Because of Neumanns principle which states that "The symmetry elements of any physical crystal property must include the symmetry elements of the crystal point group" an agreement with group theory would be a solid confirmation of the validity of SBHM. Such question has been answered by us in Ref. [20] where we have showed that this is indeed the case especially if Kleinman symmetry is valid. Kleinman symmetry can be loosely defined as the condition where one can freely permute all the indices in the susceptibility. For SHG only one input frequency is used therefore Kleinman symmetry requires:

$$\chi^{(2)}_{jkl}(2\omega = \omega + \omega) = \chi^{(2)}_{jlk}(2\omega = \omega + \omega) = \chi^{(2)}_{kjl}(\omega = 2\omega - \omega) =$$
$$\chi^{(2)}_{klj}(\omega = -\omega + 2\omega) = \chi^{(2)}_{ljk}(\omega = 2\omega - \omega) = \chi^{(2)}_{lkj}(\omega = -\omega + 2\omega)$$
(6)

This kind of symmetry generally holds if the frequency of the laser is far below the resonance frequency $\omega_0$. Fortunately, many optical processes in material characterization applies a frequency bandwith that is far below the material resonant frequency.

The far field SHG intensity in SBHM is given by [11,12]:

$$\vec{E}_{ff}(\vec{r}) = \frac{k^2 e^{ikr}}{4\pi\varepsilon_0 r}\left[(I - \hat{k}\hat{k}) \cdot \vec{P}\right]$$
(7)

where $\hat{k}$ is the direction of the outgoing SHG wave vector and $\vec{P}$ is the SHG polarization..

The intensity is obtained by multiplying Eq. (7) with its complex conjugate partner. Fig. 4 (*left*) is a demonstration of the bond model ability in correctly predicting RAS experimental data of the SHG intensity as a function of z-axis rotation for a Si(111) facet. A femtosecond laser and rotational anisotropy spectroscopy (Fig. 5 *right*) were operated at the Christian Doppler Laboratory - MACH ultrashort laser lab at the Johannes Kepler University in Linz, Austria to obtain the experimental data and a more detailed description can be found in Ref. [12].

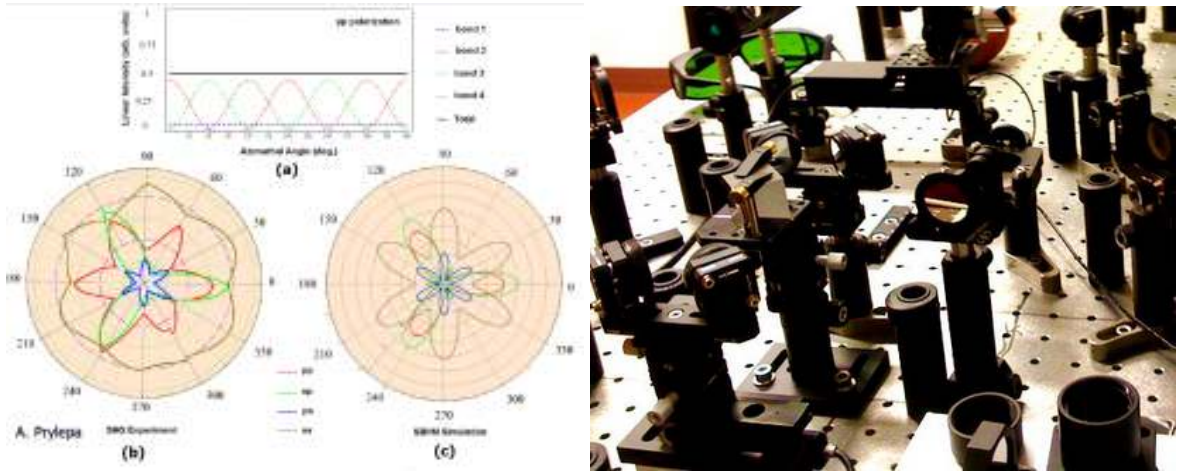

Figure 5. Experimental data and SBHM simulation of a Si111 facet for the linear (top left) and SHG (top down) intensity as a function of the azimutha rotation angle. Experimental Setup (right).

The SBHM polarization when evaluated for the linear susceptibility given by:

$$\vec{P}_1 = \frac{1}{V}\sum_{j=1}^{4}\alpha_{1j}b_j(\phi)b_j(\phi)\cdot\vec{E}$$
(8)

shows an isotropic behaviour: it is independent of the bond rotation along the z-axis. However when evaluated for the nonlinear susceptibility given by

$$\vec{P}_2 = \frac{1}{V}\sum_{j=1}^{4}\alpha_{2j}b_j(\phi)\left(b_j(\phi)\cdot\vec{E}\right)^2$$
(9)

is anisotropic as the intensity depends on the azimuth angle $\phi$. For example, the s-in s-out polarization case far field is in the form of

$$E_{2\omega,ss} = -\frac{3}{4}\alpha_d \sin^3\beta \sin 3\phi \hat{y} \qquad (10)$$

here $\beta$ is the angle between each of the Si bond and $\alpha_d$ is the hyperpolarizability of the bond. Therefore more information can be extracted from nonlinear susceptibilities and this is one reason why the nonlinear susceptibility is applied in this case to study the surface properties rather than the linier susceptibility.

## 4. THE FUTURE OF SBHM

Because of its clear physical picture, the model offers many promising application in the future. One of them is the investigation of molecular bond orientation in the interface or surface-bulk boundary. It is also possible although more complicated due to the increase in the number of independent parameters  to study surface reconstruction using SBHM because it is possible to determine is specific cases wheter the signal comes from the interface or from the bulk, the later whose point group is usually known. Using real time monitoring it is also possible to see the bond orientation during a crystal surface growth e.g. by deposition of oxyde layers or to detect the formation of new surface molecules when attached by the environment. Thus monitoring real time kinetics of biosensors by measuring the nonlinear intensity before, during, and after molecular attachment from the environment to the biosensor sensitive surface may be possible. By looking at the changes in the intensity signal and peaks one might extract useful information about surface processes.

Applying more than one input laser field e.g. Sum Frequency Generation (SFG) is also possible to access different energy levels to study the bandgap of semiconductors such as surface states energy bandgaps. By using one laser pump at constant frequency  and one as the varying frequency input one can study Raman active material. If the intensity of the even harmonic signal is low then the energy is not sufficient to excite an electron across the surface bandgap and vice versa.  Moreover, Spatial dispersion, bulk quadrupole, and magnetic terms may also be added in the model to study more complex thin film materials

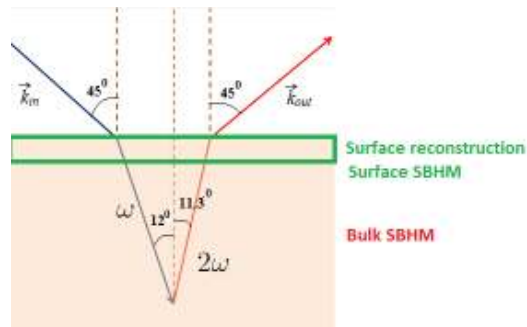

Figure 6.  Molecular bond orientation at the surface can be different with inside the bulk, this can also be applied for monitoring surface orientation for biosensors.

## 5. SUMMARY

It has been briefly demonstrated that the bond model is a nonlinear nano optical phenomenolgical method which can be applied to study semiconductor surfaces that posesses parity symmetry. By assuming motion only along the bond it is possible to explore the surface orientation of a material in real time and udner room condition. Many interesting application using the bond model can be expected in the future such as studying surface orientation and material interface band gaps levels.

## 6. ACKNOWLEDMENT

I would like to thank Prof. Hingerl and Dr. Alejo-Molina for previous fruitful discussions regarding the bond model and crystal symmetry. I also am grateful to Andrii Prylepa for providing the Figures of the experimental femtosecond setup at the CDS-MACH Laboratory of ZONA JKU Linz..